\documentclass[letterpaper, 10 pt, conference]{ieeeconf}
\IEEEoverridecommandlockouts
\usepackage{amsmath,amsfonts,amssymb}

\DeclareMathOperator*{\argmin}{arg\,min}
\usepackage{mathtools}
\usepackage{algorithmic}
\usepackage{algorithm}
\usepackage{array}
\usepackage{textcomp}
\usepackage{stfloats}
\usepackage{url}
\usepackage{verbatim}
\usepackage{multirow}
\usepackage{graphicx}
\usepackage{cite}
\usepackage[capitalise]{cleveref}
\usepackage{epsfig}
\usepackage{textcomp}
\usepackage{xcolor}
\usepackage{cases}
\usepackage{subcaption}
\graphicspath{{Images/}}
\title{\LARGE \bf
A Continuification Approach to CAV Control in Mixed Traffic via Variable Speed Limits
}

\author{B. Block$^{1}$, C. Pasquale$^{2}$, S. Siri$^{2}$, S. Sacone$^{2}$, S. Stockar$^{1}$
\thanks{This material is based upon work supported by the National Science Foundation CAREER Award 2042354.}
\thanks{$^{1}$B. Block and S. Stockar are with the Department of Mechanical and Aerospace Engineering and the Center for Automotive Research, The Ohio State University, 930 Kinnear Road, Columbus, OH 43212 USA.}%
\thanks{$^{2}$C. Pasquale, S. Siri, and S. Sacone are with the Department of Informatics, Bioengineering, Robotics, and Systems Engineering, University of Genova, Italy.}%
}
\begin{document}

\maketitle
\thispagestyle{empty}
\pagestyle{empty}

\begin{abstract}

This paper presents a method for controlling traffic via the use of connected and automated vehicles (CAVs) acting as moving bottlenecks. Current methods for moving bottleneck control use a coupled PDE-ODE model, based on the Lighthill-Whitham-Richards (LWR) model, to represent the influence of the CAV. Control of the CAV is normally achieved by designing the control on the ODE which models the speed of the moving bottleneck. The proposed method in this paper instead looks to reduce the computational burden of controlling multiple CAVs by designing the moving bottleneck controller first on the PDE. The original control designed on the PDE is a linear quadratic regulator (LQR) that determines the optimal variable speed limit (VSL) for the entire length of freeway in order to regulate density to a desired setpoint. Then, a \textit{continuification} approach is utilized to determine the input speed for each CAV.  Results show that multiple CAVs can be controlled via this method, with minimal computational burden, and that as the number of CAVs increases the solution approaches the global optimal solution determined by the LQR.

\end{abstract}

\section{INTRODUCTION}
Reducing congestion in traffic and regulating the system toward a desired density profile is the focus of many traffic control strategies \cite{block_lq_2024,siri_freeway_2021,yu_reinforcement_2022}. Control strategies for traffic flow can include ramp metering \cite{yu_reinforcement_2022}, variable speed limit (VSL) control \cite{block_lq_2024}, or the use of connected and automated vehicles (CAVs) as control actuators \cite{block_analysis_2025,daini_traffic_2025}.

Commonly for control, traffic is modeled via partial differential equations (PDE) that describe the density and flow of vehicles through a system \cite{siri_freeway_2021}. These PDE models are preferable as they are able to accurately model traffic while being computationally efficient. Control techniques designed on these PDE models are generally either boundary ramp metering control \cite{yu_reinforcement_2022} or infinite-dimensional, in-domain VSL control \cite{karafyllis_feedback_2019-1,block_lq_2024,block_lq-informed_2025}. Recently, though, using CAVs to control traffic flow has gained prominence \cite{delle_monache_scalar_2014,piacentini_traffic_2021,daini_traffic_2025,block_analysis_2025}.

The influence of CAVs on the bulk traffic flow is modeled via an ordinary differential equation (ODE) flow constraint called a moving bottleneck \cite{block_analysis_2025}. This creates a coupled PDE-ODE model where the control actuator is the speed of the moving bottleneck. Control of these coupled PDE-ODE models is normally applied directly on the ODE \cite{piacentini_traffic_2021,daini_traffic_2025,piacentini_multiple_2019,liard_optimal_2020}. In \cite{piacentini_multiple_2019} a proportional-integral (PI) controller was developed to control multiple vehicles in a platoon to reduce congestion. In both \cite{piacentini_traffic_2021,daini_traffic_2025}, model predictive control (MPC) was used to control multiple vehicles or a platoon to reduce traffic energy consumption. An optimal control algorithm was developed for CAV control in \cite{liard_optimal_2020} in order to reduce energy as well. All of these methods were designed directly on the ODE constraint, so when the number of CAVs increases, so does the computation time \cite{daini_traffic_2025}.

An alternative approach is to design the control directly on the PDE. In multi-agent system control, this process is called \textit{continuation} \cite{nikitin_continuation_2022,Fueyo_Continuation_2025} or \textit{continuification} \cite{maffettone_continuification_2023-1,Maffettone_Leader_2025}. This paper will use the term \textit{continuification}. In this process, a control strategy is designed at the macroscopic level, derived from the agents' dynamics and then it is discretized to be used by the agents at a microscopic level. In \cite{nikitin_continuation_2022} this process was used to control a swarm of agents to follow a specific profile. In both \cite{maffettone_continuification_2023,Maffettone_Leader_2025} it was used to control agents to form a desired profile on a ring. The process was used in \cite{Fueyo_Continuation_2025} to stabilize traffic on a circular road. In all these approaches, though, the control was applied to every agent, or vehicle, in the system, which is not always possible.

A key limitation of the current literature on CAV control for traffic flow is that the control is developed directly on the ODE. This becomes computationally expensive when multiple CAVs or platoons are considered. In contrast, when PDE control of multi-agent systems is considered, every agent in the system is considered controllable which is not always a valid assumption. This paper aims to address these limitations by designing a PDE-based controller for a sparse amount of CAVs in traffic flow in order to reduce congestion. Starting from the underlying macroscopic PDE traffic model, an optimal, infinite-dimensional VSL controller previously derived by the authors \cite{block_lq_2024} is used. Then, the VSL control action is discretized to individual CAVs in traffic flow via \textit{continuification} in order to drive traffic density to a desired profile. To show the efficacy of the proposed method, the PDE-based CAV controller is compared against an ODE-based MPC for the moving bottleneck model. The novel PDE-based approach is shown to be both more efficient in reaching the desired density profile and is shown to approach the inifite-dimensional global optimal solution while being more computationally efficient than the ODE-based approach.

This paper is structured as follows. \cref{sec:Models} presents the macroscopic PDE and coupled PDE-ODE model used in the moving bottleneck approach. \cref{sec:Controllers} presents the different control architectures, namely the previously derived infinite-dimensional VSL controller, the novel PDE-based moving bottleneck controller, and the baseline ODE-based moving bottleneck controller. The performance of the controllers is analyzed in \cref{sec:Results} and concluding remarks are given in \cref{sec:Conclusion}.

\section{TRAFFIC MODELS}\label{sec:Models}

\subsection{Macroscopic PDE Traffic Model}
The macroscopic traffic model used for both the infinite-dimensional control and the moving bottleneck approach is the Lighthill-Whitham-Richards (LWR) model \cite{lighthill_kinematic_1955,richards_shock_1956} which is a first-order scalar hyperbolic PDE given by
\begin{equation} \label{eq:LWR}
    \frac{\partial\rho(z,t)}{\partial t} + \frac{\partial f(\rho(z,t))}{\partial z} = 0
\end{equation}
where $\rho(z,t)$ is the total traffic density, $z$ is the spatial variable, and $t$ is time. As the LWR model is first-order it relies on an equilibrium relation, or fundamental diagram, $f(\rho(z,t))$, to describe the vehicle flow. 

\subsection{Coupled PDE-ODE Moving Bottleneck Model}
To model the effect that one or more CAVs have on the traffic flow, an inequality flow constraint is added to the model, creating a coupled PDE-ODE framework. The position of the inequality constraint in the flow is at the position(s) of the CAV(s). The complete coupled PDE-ODE model is given as
\begin{subnumcases}{}\label{eq:coupledPDEODE}
   \label{eq:coupledPDEODE_PDE}\frac{\partial\rho(z,t)}{\partial t} + \frac{\partial f(\rho(z,t))}{\partial z} = 0 \\
   \label{eq:coupledPDEODE_PDE_IC}\rho(z,0) = \rho_0(z) \\
   \label{eq:coupledPDEODE_PDE_flux}f(\rho(y(t),t)) - \Dot{y}(t)\rho(y(t),t) \leq F_\alpha(\rho_\alpha) \\
   \label{eq:coupledPDEODE_ODE}\Dot{y}(t) = \omega(\rho(y(t)+,t)) \\
   \label{eq:coupledPDEODE_ODE_IC}y(0) = y_0 
\end{subnumcases}
where $\rho_0(z)$ is the initial density profile, $y(t)$ is the position of the CAV with $y_0$ being its initial position, and $\Dot{y}(t)$ is the CAV's velocity which is determined by
\begin{equation}
    \omega(\rho(z,t)) = 
    \begin{cases}
        V_\mathrm{AV} & \rho(z,t)\leq\rho^*\\
        v(\rho(z,t)) & \textrm{otherwise}
    \end{cases}
\end{equation}
where $v(\rho(z,t))$ is the speed of surrounding traffic and $V_\mathrm{AV}$ is the control parameter. The CAV can travel at its desired speed $V_\mathrm{AV}$ if traffic is not too congested, but when the density of the bulk traffic flow around the vehicle is too high, in this case the right-hand limit at $y(t)$, the CAV changes its speed to adapt to the surrounding traffic. 

The right-hand side of \cref{eq:coupledPDEODE_PDE_flux} is derived by looking at the problem from the reference frame of the CAV \cite{delle_monache_scalar_2014} meaning
\begin{equation} \label{eq:CAVrefFlux}
    F(\rho(z,t)) = f(\rho(z,t)) - \Dot{y}(t)\rho(z,t)
\end{equation}
and
\begin{equation}
    F_\alpha(\rho_{\alpha}) = f_\alpha(\rho_\alpha) - \Dot{y}(t)\rho_\alpha
\end{equation}
where $\alpha\in[0,1]$ is a dimensionless reduction rate of the road capacity. The value of $\alpha$ is a function of the number of lanes that a bottleneck occupies given by
\begin{equation}\label{eq:alpha}
    \alpha = 1 - \frac{\#\;\mathrm{lanes\;occupied}}{\mathrm{total}\;\#\;\mathrm{of\;lanes}}
\end{equation}
The subscript $\alpha$ denotes the rescaled flow value due to the restriction.

\begin{figure}
    \centering
    \includegraphics[width=\columnwidth]{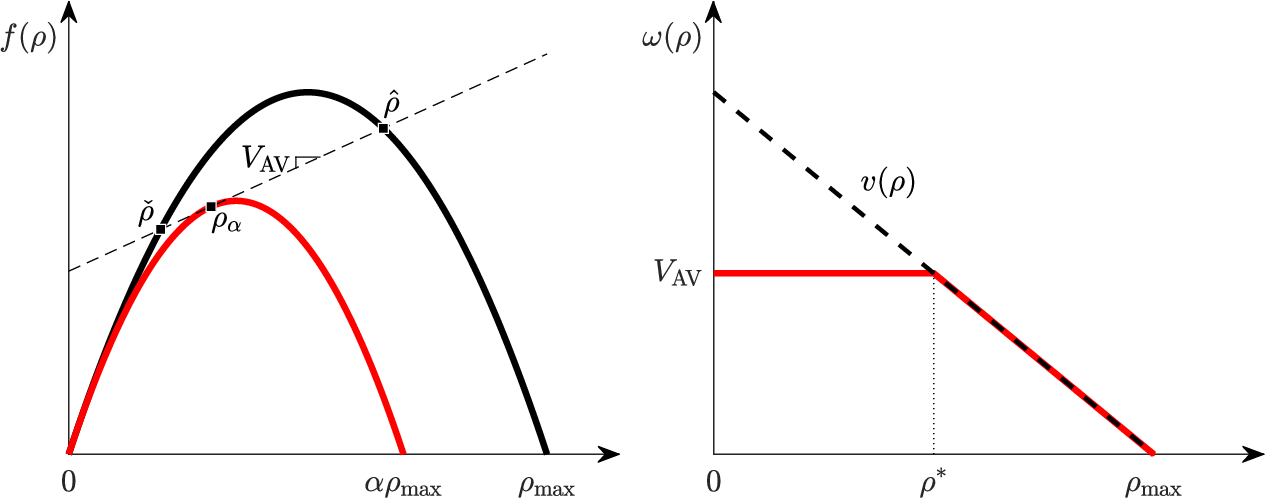}
    \caption{Reduced flow effect due the the presence of a CAV (left) and the velocity function for the CAV (right).}
    \label{fig:fundamentaldiagrams}
\end{figure}

In both the general PDE model and the coupled PDE-ODE model, the Greenshield fundamental diagram
\begin{equation}\label{eq:Greenshield}
    f(\rho(z,t))=\rho(z,t)V_\mathrm{max}\bigg(1-\frac{\rho(z,t)}{\rho_\mathrm{max}}\bigg)
\end{equation}
is used, where $V_\mathrm{max}$ is the maximum speed and $\rho_\mathrm{max}$ is the jam density. The reduced fundamental diagram is given as
\begin{equation} \label{eq:reducedgreenshieldflux}
    f_\alpha(\rho(z,t)) = \rho(z,t) V_\mathrm{max}\bigg(1 - \frac{\rho(z,t)}{\alpha\rho_\mathrm{max}}\bigg)
\end{equation}
The value $\rho_\alpha\in\big[0,\frac{\alpha\rho_\mathrm{max}}{2}\big]$ is found by
\begin{equation} \label{eq:fprimeydot}
    f'_\alpha(\rho_\alpha)=\Dot{y}
\end{equation}
which can be solved, resulting in
\begin{equation} \label{eq:rho_alpha}
    \rho_\alpha = \frac{\alpha\rho_\mathrm{max}}{2}\bigg(1 - \frac{\Dot{y}}{V_\mathrm{max}}\bigg)
\end{equation}
The upstream density $\Hat{\rho}$ and downstream density $\Check{\rho}$ around the moving bottleneck can then be found as the intersection of the original flow function $f(\rho)$ with the line $f_\alpha(\rho_\alpha) + \Dot{y}(\rho-\rho_\alpha)$ \cite{delle_monache_scalar_2014}. The impact of a moving bottleneck on the fundamental diagram is shown in \cref{fig:fundamentaldiagrams}, where the intersections $\Hat{\rho}$, $\Check{\rho}$, and $\rho_\alpha$ are shown as well as the point $\rho^*$, the density at which the CAV becomes uncontrollable as it must follow the speed of surrounding traffic.

\section{CONTROL FORMULATION}\label{sec:Controllers}

\subsection{Inifinite-Dimensional Optimal Controller}
Here, a continuous, optimal VSL controller is developed on the PDE for the basis of the optimal CAV controller for the moving bottleneck model. The infinite-dimensional linear quadratic controller described in this section was previously derived in \cite{block_lq_2024}. The cost functional to be minimized by the control input $u_\mathrm{opt}$ is given as
\begin{equation} \label{eq:costfunct}
    J(x_0,u) = \int_{0}^{\infty} \langle x(t),Qx(t) \rangle + \langle u(t), Ru(t) \rangle \,dt
\end{equation}
The solution of \cref{eq:costfunct} can be found by finding the non-negative self-adjoint operator $P$ which solves
\begin{equation} \label{eq:ORE}
    [A^*P + PA + Q - PBR^{-1}B^*P]x = 0
\end{equation}
When $(A,B)$ is exponentially stabilizable and $(C,A)$ is exponentially detectable \cref{eq:ORE} has a unique, non-negative solution $P$ and \cref{eq:costfunct} is minimized by the unique control input $u_\mathrm{opt}$ given by
\begin{equation} \label{eq:u_opt1}
    u_\mathrm{opt}(z,t)=-R^{-1}B^*Px(z,t)
\end{equation}
The control input $u_\mathrm{opt}$ for the controller is the continuous change of the speed limit over the stretch of road considered. 

The LWR model \cref{eq:LWR} with the Greenshield fundamental diagram \cref{eq:Greenshield} can be linearized and written in the form
\begin{equation} \label{eq:standardlinearPDE}
    \begin{aligned}
        \frac{\partial x}{\partial t}(z,t) &= \mathcal{V}\frac{\partial x}{\partial z}(z,t) + Mx(z,t) + Bu(z,t) \\
        y(z,t) &= Cx(z,t)
    \end{aligned}
\end{equation}
where it is linearized around a desired density profile $\rho_\mathrm{des}$ such that
\begin{equation}
    x(z,t)\coloneqq\Delta\rho=\rho-\rho_\mathrm{des}
\end{equation}
The goal of the infinite-dimensional VSL controller is to drive the system to the desired density $\rho_\mathrm{des}$. \cref{eq:standardlinearPDE} has an equivalent state space formulation on the Hilbert space $\mathcal{H}$ \cite{aksikas_lq_2009} given as
\begin{align} \label{eq:pdeSS}
    \dot{x}(t) &= Ax(t) + Bu(t) \\
    y(t) &= Cx(t)
\end{align}
where $A$ is a linear operator defined as
\begin{equation} \label{eq:A_def}
    A = \mathcal{V}\cdot\frac{d.}{dz} + M\cdot I
\end{equation}
on the domain
\begin{equation} \label{eq:domain}
    D(A) = \{x\in\mathcal{H}:\frac{dx}{dz}\in\mathcal{H}\textrm{ and }x(0)=0\}
\end{equation}
with $\mathcal{V}\in\mathbb{R}^{n\times n}$ being a symmetric matrix and $M$, $B$, and $C$ being real continuous space-varying matrix functions. It is proven in \cite{aksikas_lq_2009} that if $\mathcal{V}<0$ then $A$ generates an exponentially stable $C$-semigroup. For the case when $\mathcal{V}>0$, as it is when $\rho>\frac{\rho_\mathrm{max}}{2}$, the uniqueness of the solution to \cref{eq:ORE} can be proven by defining the operator $A$ on the domain
\begin{equation} \label{eq:domain2}
    D(A) = \{x\in\mathcal{H}:\frac{dx}{dz}\in\mathcal{H}\textrm{ and }x(1)=0\}
\end{equation}
The operator $A$ is thus exponentially stabilizable on this domain. Following \cite{block_lq_2024,block_lq-informed_2025} the solution to \cref{eq:ORE} can be determined analytically as
\begin{equation}\label{eq:P}
    P(z) = \frac{\sqrt{QR}\bigg(\exp\bigg(\frac{2B\sqrt{Q}}{\mathcal{V}\sqrt{R}}(z-c)\bigg)-1\bigg)}{B\bigg(\exp\bigg(\frac{2B\sqrt{Q}}{\mathcal{V}\sqrt{R}}(z-c)\bigg)+1\bigg)}
\end{equation}
where
\begin{equation}
    c = \begin{cases}
        L, & \rho\leq\frac{\rho_\mathrm{max}}{2} \\
        0, & \rho>\frac{\rho_\mathrm{max}}{2}
        \end{cases}
\end{equation}
Thus, by inserting \cref{eq:P} into \cref{eq:u_opt1} and choosing $Q$ and $R$, the variable speed limit across the entire length of road that drives the density to its desired value can be obtained. The next section uses this speed limit to assign desired speeds to individual CAVs using the coupled PDE-ODE framework given in \cref{eq:coupledPDEODE_PDE,eq:coupledPDEODE_PDE_IC,eq:coupledPDEODE_PDE_flux,eq:coupledPDEODE_ODE,eq:coupledPDEODE_ODE_IC}.

\subsection{PDE-Based Finite-Dimensional CAV Controller}

\begin{figure}
    \centering
    \includegraphics[width=\columnwidth]{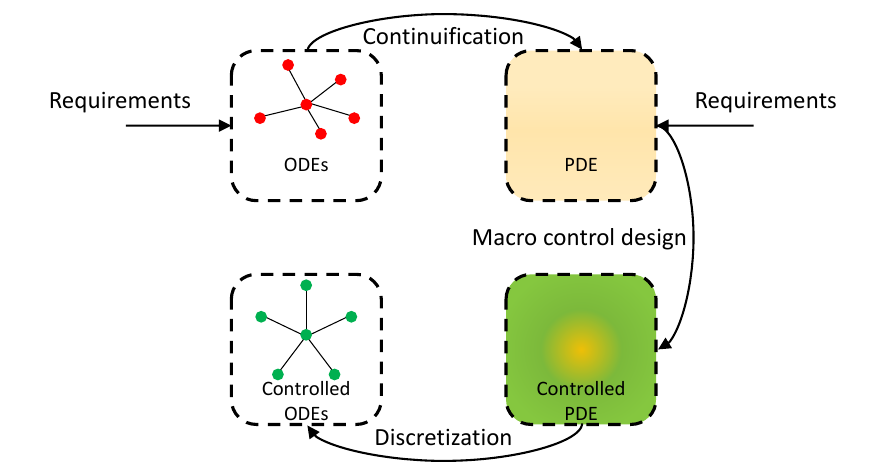}
    \caption{The process of \textit{continuification} adapted from \cite{maffettone_continuification_2023}.}
    \label{fig:continuificationdiagram}
\end{figure}

In this section, the goal is to use individual CAVs $i$ to control traffic and drive the system to a desired density profile $\rho_\mathrm{des}$. The inifinite-dimensional VSL control derived previously will be used as a computationally efficient way to compute the desired CAV velocities. The method used to convert the infinite-dimensional VSL control input into a finite-dimensional CAV speed controller is inspired by the theory of \textit{continuification} \cite{maffettone_continuification_2023}. The process of using \textit{continuification} for control design is showcased in \cref{fig:continuificationdiagram}. Starting from a system of coupled ODEs, an equivalent PDE description can be obtained via conservation laws. Control design is then performed on the infinite-dimentional plant and mapped back to agents \cite{maffettone_continuification_2023,maffettone_continuification_2023-1,nikitin_continuation_2022}. Here, instead of starting from a system of ODEs, the PDE system is immediately defined as it has been shown that PDE traffic representations are equivalent to their ODE counterparts \cite{block_physics-inspired_2024,block_analysis_2025}. The infinite-dimensional controller was defined in the previous section. The output of the PDE control design in the case of VSL control is the ratio of the speed limit or more concretely
\begin{equation}\label{eq:completespeedcontrol}
    V_\mathrm{VSL} = V_\mathrm{max,0}\int_0^L u_\mathrm{opt}(z,t)\:dz
\end{equation}
where $u_\mathrm{opt}(z,t)$ is the control found by solving \cref{eq:ORE,eq:u_opt1}. Then, the control action is discretized back to the multi-agent system by applying the control input only at the location of a controllable agent or
\begin{equation}\label{eq:CAVu_opt}
    u_i(t) = U(z_i,t) = V_\mathrm{VSL}(z_i,t)
\end{equation}
where $u_i(t)$ is the individual control action sent to agent $i$ and $z_i$ is the location of that agent.

Thus, the controller affects the speed of the moving bottleneck $\omega$ as
\begin{equation}\label{eq:controlomega}
    \omega(\rho(z,t)) = \begin{cases}
        u_i(t), & \rho\leq\rho^*\\
        v(\rho(z,t)), & \mathrm{otherwise}
    \end{cases}
\end{equation}
The algorithm for computing the control input is shown in \cref{alg:Continuation}. Unlike previous continuification problems, here not every vehicle is being controlled, so there is still a bulk flow that is being described by the PDE, \cref{eq:LWR}. Because of this, there are instances where the system can become uncontrollable as the commanded velocity of the CAV is higher than the speed of traffic surrounding it and thus it must travel at the speed of traffic. This phenomenon is evaluated later in \cref{sec:Results}.

\begin{algorithm}[!t] 
    \caption{PDE-Based CAV Controller} 
	\renewcommand{\algorithmicrequire}{\textbf{Input:}}
	\renewcommand{\algorithmicensure}{\textbf{Output:}}
    \begin{algorithmic}[1] 
    	\REQUIRE
    		Traffic density $\rho$; LQR parameters $Q$, $R$; Initial position of CAV $y_0$; Initial speed of CAV $\Dot{y}_0$; lane occupancy $\alpha$. 
    	\ENSURE
    		CAV optimal speed $u_i(t)$.
        \WHILE{$t \leq T_f$}
            \STATE
                Solve \cref{eq:coupledPDEODE_PDE} and compute linearized model.
            \STATE
        	    Compute state feedback function \cref{eq:P} for entire stretch of road.
            \STATE
                Calculate the position $y_i(t)$ of each CAV $i$.
            \STATE
                Assign CAV $i$ the speed determined by \cref{eq:u_opt1,eq:completespeedcontrol} at position $y_i$ as in \cref{eq:CAVu_opt}.
            \STATE
                Solve the coupled PDE-ODE model \cref{eq:coupledPDEODE_PDE,eq:coupledPDEODE_PDE_IC,eq:coupledPDEODE_PDE_flux,eq:coupledPDEODE_ODE,eq:coupledPDEODE_ODE_IC}.  
        \ENDWHILE
        \STATE $\Dot{y}_i= u_i(t)$.
    \end{algorithmic} 
	\label{alg:Continuation} 
\end{algorithm}

\subsection{Baseline ODE-Based CAV Controller}

In order to determine the effectiveness of the newly designed CAV controller, it is compared against the current state-of-the-art. Currently, for moving bottleneck control, the controller is always designed on the ODE through either a PI-type feedback \cite{piacentini_multiple_2019} or through MPC \cite{daini_traffic_2025}. For the baseline controller considered here, a centralized MPC is chosen as its computation is based on complete information about the macroscopic traffic state as well as the speed of each controlled CAV, like the PDE-based controller.

The MPC control strategy implemented at time $t$ on a time horizon $\Delta T$ takes each CAV position $y_i(t)$ and the current traffic density $\rho(z,t)$ as inputs and computes the optimal control that minimizes a selected cost function. Like the previous controllers, the baseline MPC also aims to drive the density to a desired profile. The cost function is the root mean square error (RMSE) at each time step given by
\begin{equation}\label{eq:RMSE}
    \mathrm{RMSE}(t) = \sqrt{\frac{\sum_{z=1}^{N_z}||\rho(z,t)-\rho_\mathrm{des}(z,t)||^2}{N_z}}
\end{equation}
and the optimal velocity for each CAV is then given as
\begin{equation}\label{eq:ODEcostfunct}
    u_\mathrm{opt}(t) = \argmin \mathrm{RMSE}
\end{equation}
subject to \cref{eq:coupledPDEODE_PDE,eq:coupledPDEODE_PDE_IC,eq:coupledPDEODE_PDE_flux,eq:coupledPDEODE_ODE,eq:coupledPDEODE_ODE_IC} and constraints
\begin{equation}\label{eq:ODEconstraints}
    u_\mathrm{min}\leq u_i(t) \leq u_\mathrm{max}
\end{equation}
where $u_\mathrm{min}=0$ and $u_\mathrm{max}=v(\rho(z,t))$, or the speed of the surrounding traffic. The cost function is nonlinear and nonconvex, so the constrained optimization problem \cref{eq:ODEcostfunct,eq:ODEconstraints} is solved using the MATLAB function {\tt fmincon}, as in \cite{daini_traffic_2025}.

\section{SIMULATION RESULTS}\label{sec:Results}

\begin{figure*}[ht!]
    \centering
    \begin{subfigure}{0.24\textwidth}
            \centering
            \includegraphics[width=\textwidth]{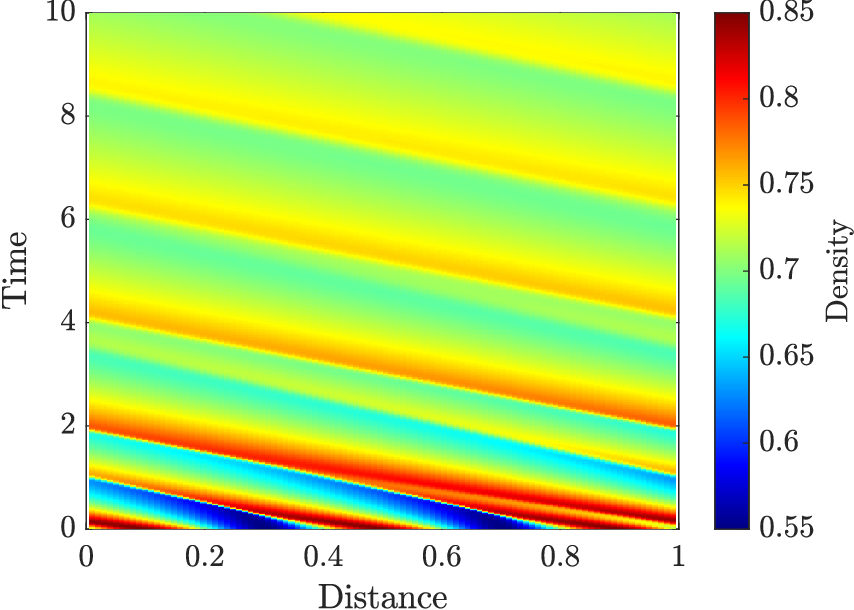}
            \caption{}
            \label{fig:Rho_unc}
    \end{subfigure}
    \begin{subfigure}{0.24\textwidth}
            \centering
            \setcounter{subfigure}{2}
            \includegraphics[width=\textwidth]{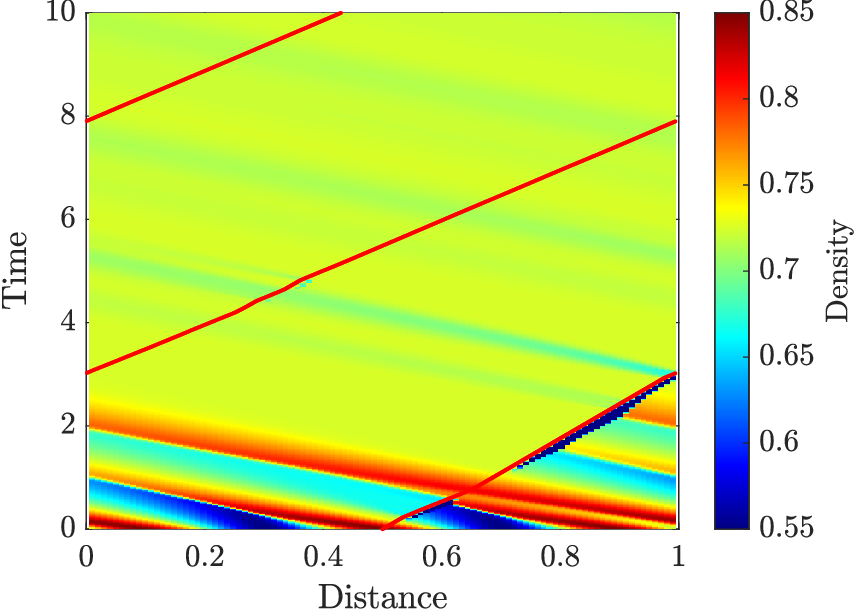}
            \caption{}
            \label{fig:Rho_cav_1veh_min}
    \end{subfigure}
    \begin{subfigure}{0.24\textwidth}
            \centering
            \setcounter{subfigure}{4}
            \includegraphics[width=\textwidth]{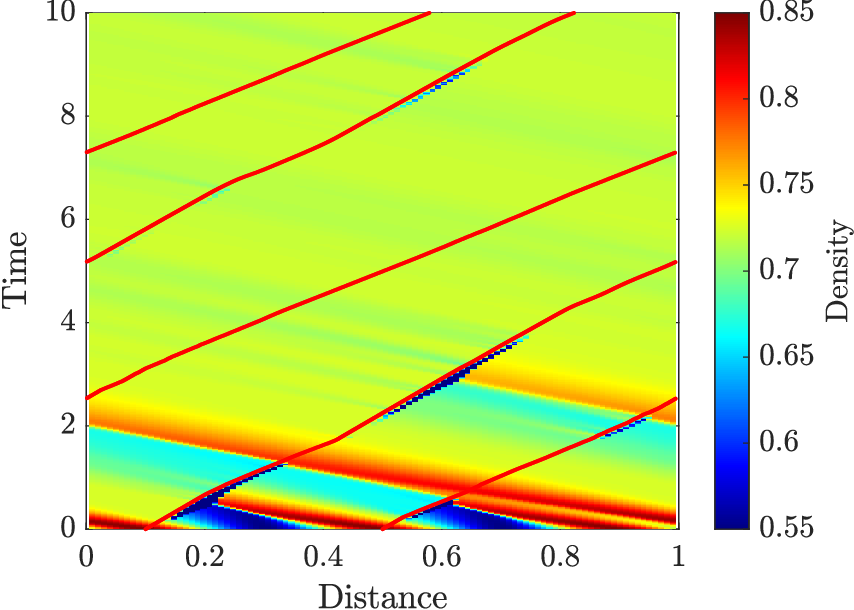}
            \caption{}
            \label{fig:Rho_cav_2veh_min}
    \end{subfigure}
    \begin{subfigure}{0.24\textwidth}
            \centering
            \setcounter{subfigure}{6}
            \includegraphics[width=\textwidth]{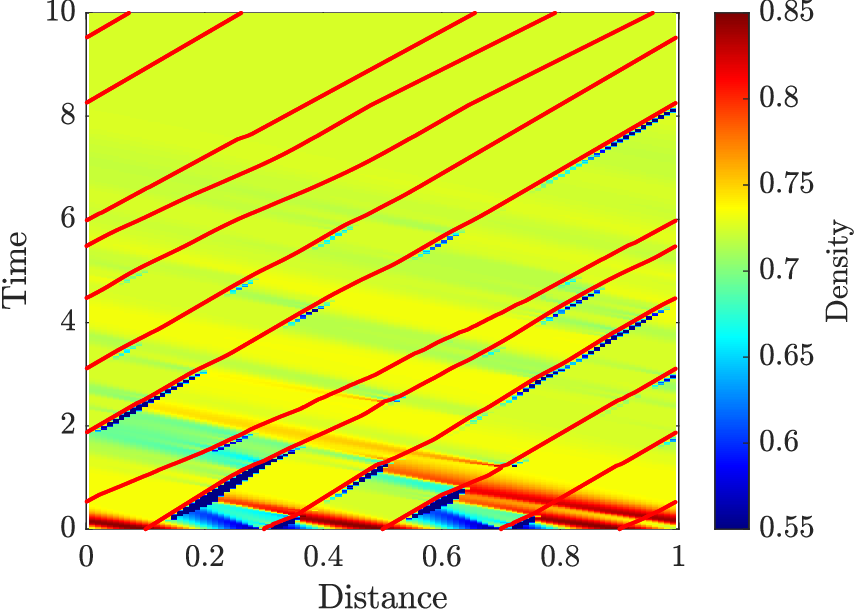}
            \caption{}
            \label{fig:Rho_cav_5veh_min}
    \end{subfigure}
    \begin{subfigure}{0.24\textwidth}
            \centering
            \setcounter{subfigure}{1}
            \includegraphics[width=\textwidth]{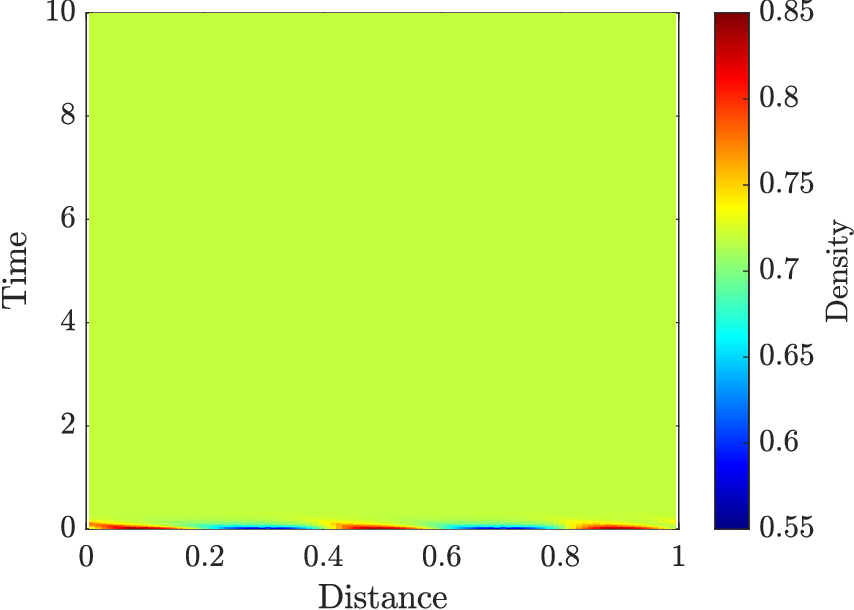}
            \caption{}
            \label{fig:Rho_lqr}
    \end{subfigure}
    \begin{subfigure}{0.24\textwidth}
            \centering
            \setcounter{subfigure}{3}
            \includegraphics[width=\textwidth]{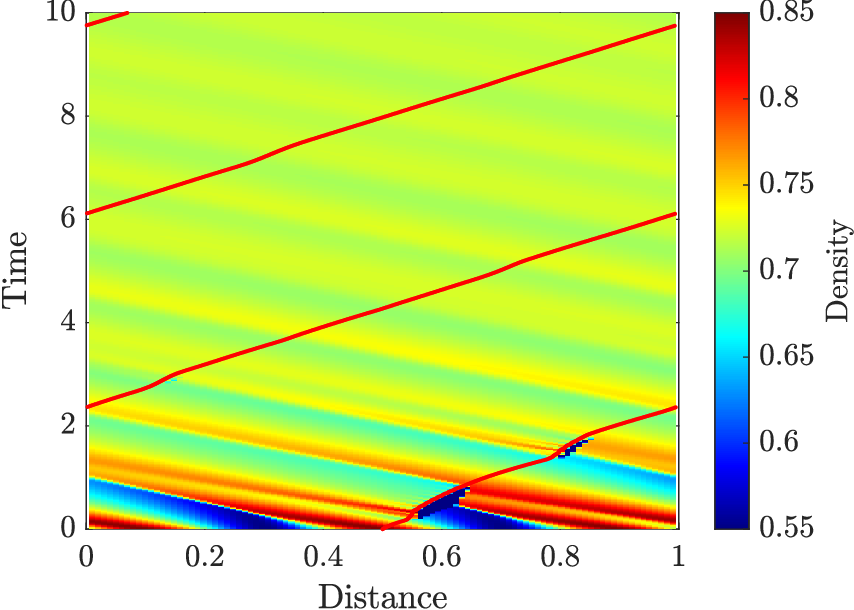}
            \caption{}
            \label{fig:Rho_cav_1veh_lqr}
    \end{subfigure}
    \begin{subfigure}{0.24\textwidth}
            \centering
            \setcounter{subfigure}{5}
            \includegraphics[width=\textwidth]{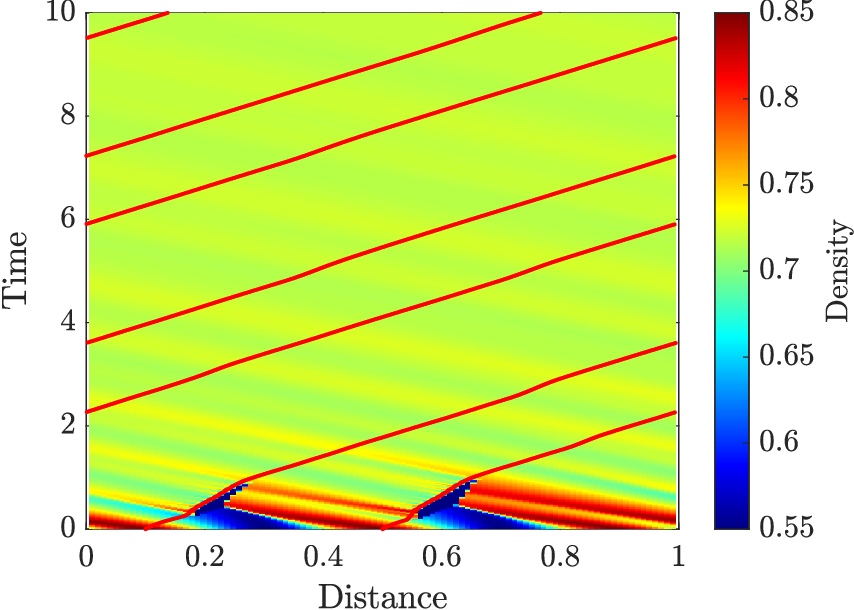}
            \caption{}
            \label{fig:Rho_cav_2veh_lqr}
    \end{subfigure}
    \begin{subfigure}{0.24\textwidth}
            \centering
            \setcounter{subfigure}{7}
            \includegraphics[width=\textwidth]{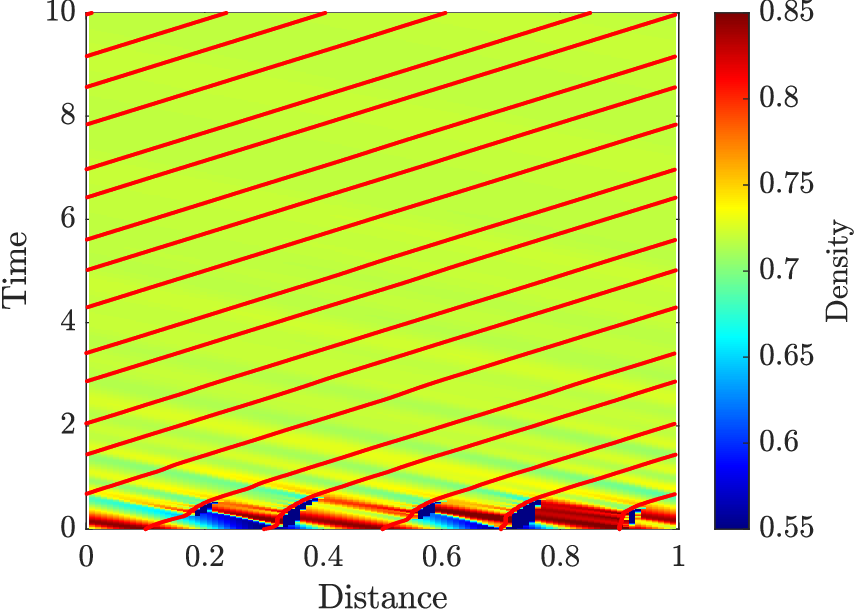}
            \caption{}
            \label{fig:Rho_cav_5veh_lqr}
    \end{subfigure}
    \caption{Traffic density for (a) the uncontrolled case, (b) the infinite-dimensional global optimal VSL case, (c) one CAV controlled via ODE-based MPC, (d) one CAV controlled via PDE-based control, (e) two CAVs controlled via ODE-based MPC, (f) two CAVs controlled via PDE-based control, (g) five CAVs controlled via ODE-based MPC, (h) five CAVs controlled via PDE-based control.}
    \label{fig:baseline}
\end{figure*}

To investigate the performance of each controller, a simple numerical example of a circular road is considered. The maximum density and maximum velocity, or speed limit, are both normalized to 1, i.e. $\rho_\mathrm{max}=V_\mathrm{max}=1$. The initial condition considered is
\begin{equation}\label{eq:IC}
    \rho_0(z) = 0.7 + 0.15\sin(5\pi z)
\end{equation}
with periodic boundary conditions
\begin{equation}
    \rho(0,t) = \rho(L,t)
\end{equation}
where $L$ is the normalized length of road considered, $L=1$. A two lane road is considered, so the occupancy parameter for a CAV $i$ is $\alpha=0.5$. The parameters for the LQR are $Q=1000$ and $R=0.2$. The simulation time is $T_f = 10$ time units. For both the LQR and the PDE-based CAV controller, the optimal control input is determined at each time step $t$. For the baseline ODE-based MPC CAV controller, it is implemented in a receding horizon approach meaning that $\Delta T = T_f - t$, and it is updated every time step $t$. Three different CAV cases are examined, using 1 CAV to control flow, using 2 CAVs, and using 5 CAVs.

The uncontrolled traffic density is shown in \cref{fig:Rho_unc} where it can be seen that the high density regions propagate backwards and remain in the simulation until the final time. In \cref{fig:Rho_lqr}, the global optimal infinite-dimensional VSL control is applied. It mitigates the density oscillations almost instantly and drives the traffic to the desired density $\rho_\mathrm{des}$. This can be seen as an ideal theoretical upper-limit benchmark on performance. In practical implementations, full compliance of all vehicles with VSL commands is rarely achieved. But, when speed control is actuated via CAVs, vehicles traveling in the vicinity of the controlled vehicles are effectively constrained to follow their speed, allowing for more realistic results and a more implementable strategy.

\begin{figure}
    \centering
    \begin{subfigure}{\columnwidth}
        \centering
        \includegraphics[width=\textwidth]{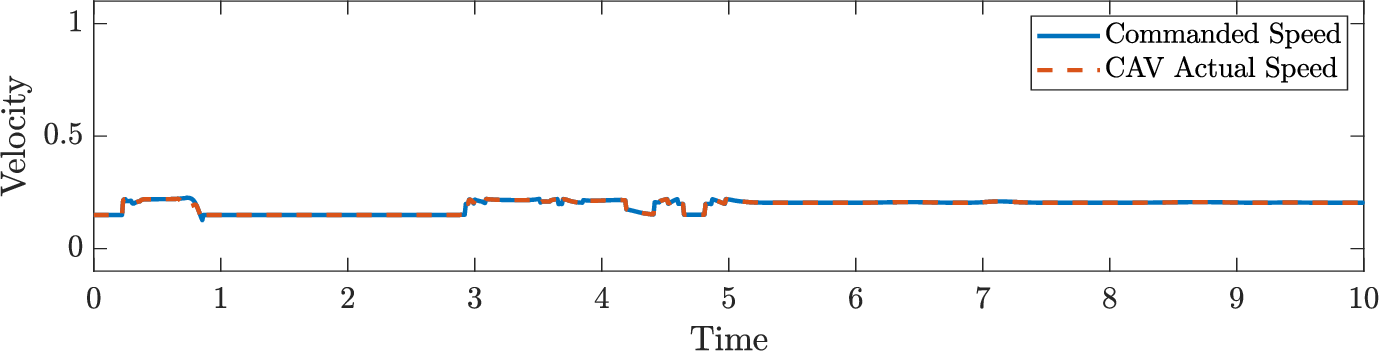}
        \caption{}
        \label{fig:Velo_min}
    \end{subfigure}
    \begin{subfigure}{\columnwidth}
        \centering
        \includegraphics[width=\textwidth]{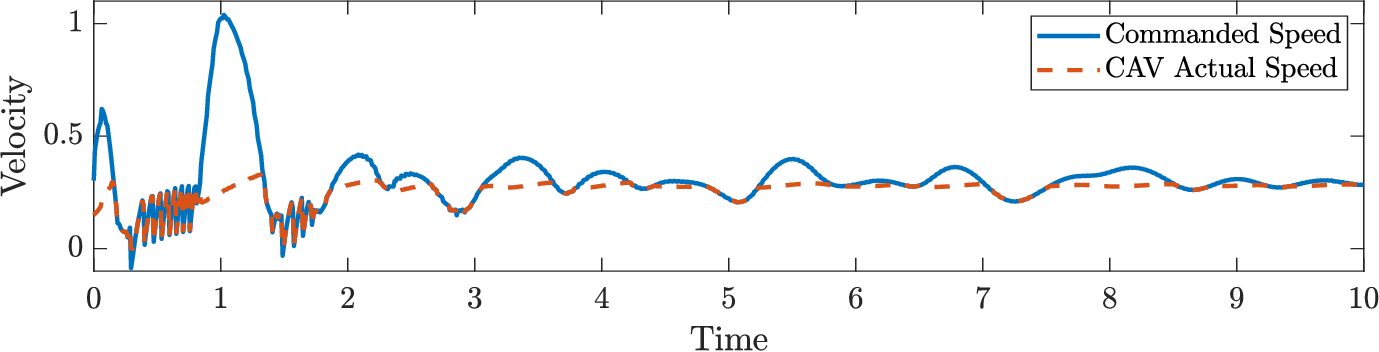}
        \caption{}
        \label{fig:Velo_lqr}
    \end{subfigure}
    \caption{Commanded and actual velocity profiles for a single CAV for (a) ODE-based MPC and (b) PDE-based controller.}
    \label{fig:velos}
\end{figure}

The single CAV case is shown in \cref{fig:Rho_cav_1veh_min} and \cref{fig:Rho_cav_1veh_lqr}. In \cref{fig:Rho_cav_1veh_min}, the controller is the ODE-based MPC and in \cref{fig:Rho_cav_1veh_lqr} the controller is the PDE-based \textit{continuification}-style controller. Both have similar results in driving the system to the desired density, though the PDE-based controller is able to attenuate the high density wave starting at $L=1$ sooner. A difference can be seen in both the commanded and actual speed of the CAVs, though, as shown in \cref{fig:velos}. In \cref{fig:Velo_min}, the MPC-commanded and actual speed of the CAV is shown and it can be seen that they are almost the same. This is because it is a constrained optimization and the upper limit $u_\mathrm{max}$ is set to be the speed of the surrounding traffic. On the other hand, the LQR control input is unconstrained so it sometimes commands speeds that are infeasible as shown in \cref{fig:Velo_lqr}. In the case of an infeasible control input from the PDE-based controller, the CAV instead travels at the speed of surrounding traffic, following \cref{eq:controlomega}. Also, since the LQR can sometimes command negative speeds, the control input to the CAV is saturated at zero. It can be seen that the PDE-based optimal controller actuates much more often.

The 2 CAV case is shown in \cref{fig:Rho_cav_2veh_min} and \cref{fig:Rho_cav_2veh_lqr} for the ODE-based and PDE-based controllers, respectively. Here there is a bigger difference in results as the ODE-based MPC has a similar result to the single CAV case whereas the PDE-based controller is able to mitigate the traffic within 2 time units. In both cases, the CAVs are able to drive the traffic to the desired density, but the ODE-based MPC controller still needs to actuate later in the simulation, around 8 time units, while at that point the PDE-based controller is not active and the CAVs are traveling at the speed of traffic.

Lastly, the 5 CAV case is shown in \cref{fig:Rho_cav_5veh_min} and \cref{fig:Rho_cav_5veh_lqr} for the ODE-based and PDE-based controllers, respectively, where it is seen that the PDE-based approach regulates the density to the desired density profile quicker than the ODE-based MPC. Also, an increase in CAVs leads to the PDE-based method approaching the LQR global optimal solution.

\begin{figure}
    \centering
    \includegraphics[width=\columnwidth]{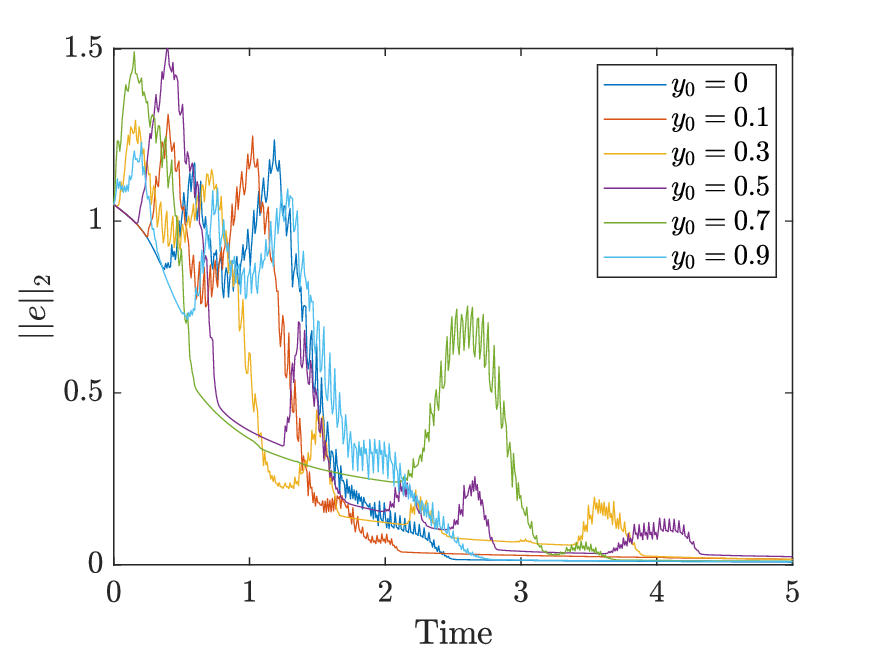}
    \caption{Density error profiles for different placements of a single CAV.}
    \label{fig:error}
\end{figure}

One aspect to note is that the initial location of the CAV $y_0$ greatly impacts the solution. This is especially evident in the single CAV case, where the control input has less control authority over the system. This is showcased in \cref{fig:error} where the $\mathcal{L}_2$ norm of the error between actual and desired density is shown for different initial starting positions of the CAV, with $y_0=0.5$ being the case shown in \cref{fig:Rho_cav_1veh_lqr}. All initial positions achieve the same result by 5 time units, but their error profiles are very different. This could be because of the surrounding density in the location of the CAV at its initial point. This is more of an issue in real traffic since with a circular road the traffic re-enters at the beginning of the road and the CAV can be placed anywhere. In real traffic, such as highways, there are only certain points at which a CAV can be inserted into the system, i.e. on-ramps. Further analysis and discussion on this is left for future work.

Besides improved density regulation performance, the PDE-based \textit{continuification} approach is much more computationally efficient. This is shown in \cref{tab:simtime} where the real time factor for each of the three CAV cases is shown. A real time factor less than 1 means that the controller is real time capable whereas a value above 1 means that it takes longer to compute than the simulation time. All simulations were run on a 1.7 GHz Intel I7 Processor with 16 GB of RAM and the times were averaged over 20 runs. As can be seen, the PDE-based approach computes quicker in every case, up to 600x faster in the 5 CAV case,  and is always real time capable. On the other hand, the ODE-based MPC is only real time capable when controlling a single CAV. This is because with each added CAV the MPC needs to compute another control input whereas the PDE-based controller always computes the optimal speed at each point via \cref{eq:P} and then assigns the commanded speed at the position of the CAV without need for more computations. One interesting thing to note is that there is not much increase in computation time for the ODE-based MPC approach going from 2 CAVs to 5 CAVs. This is because {\tt fmincon} has varying computation times depending on how quickly it can determine a minimum. It is worth noting that the computational effort required by the proposed approach remains low as the number of controlled CAVs increases. This aspect makes the proposed control approach particularly suitable for large-scale traffic networks and scenarios with increasing penetration rates of controlled CAVs. A more in-depth analysis involving large-scale networks and varying levels of penetration of controlled CAVs will be addressed in future work

\begin{table}[thpb]
    \centering
    \begin{center}
    \caption{Real time factor of the different algorithms.}
    \label{tab:simtime}
    \begin{tabular}{l|c|c|c}
    \hline \hline
    Controller    & 1 CAV & 2 CAVs & 5 CAVs\\\hline
    ODE-based MPC & 0.824 & 2.288  & 2.315\\
    PDE-based     & 0.002 & 0.003  & 0.004\\
    \hline \hline
    \end{tabular}
    \end{center}
\end{table}

\section{CONCLUSION}\label{sec:Conclusion}

This work presented a controller for CAVs in traffic flow to reduce congestion. Starting from an infinite-dimensional PDE control input, a CAV speed controller was derived through the process of \textit{continuification}. This approach was compared against an ODE-based MPC for CAV moving bottleneck control. It was shown through a simulation study that the PDE-based CAV control method performed better both in terms of density reduction and computation time. The PDE-based control was able to compute the optimal CAV trajectories in real time for all cases whereas the ODE-based controller was only able to do so for the single CAV case, with the PDE-based controller being more than 600x faster than the ODE-based and 250x faster than real time. It was also seen that as the number of CAVs increased, the controller began to approach the global optimal infinite-dimensional solution.

Future work will focus on the application of this method to more realistic traffic scenarios as well as applying the derived control actions back to a more realistic microscopic simulator as well as further analysis of how the placement of the CAVs impacts controllability.

\bibliographystyle{IEEEtran}
\bibliography{references}
\end{document}